\RequirePackage[2020-02-02]{latexrelease}
\documentclass[twocolumn,amsmath,amssymb,aps,UKenglish]{revtex4}
\usepackage{graphicx}
\usepackage{amsmath,amssymb}
\usepackage{mathrsfs}
\usepackage{color}
\usepackage{afterpage}
\usepackage[version=3]{mhchem}
\usepackage{natbib}
\usepackage{soul}
\usepackage[caption=false]{subfig}
\usepackage{array}
\usepackage{multirow}

\usepackage{isodate}
\begin{document}

\title{Accelerated Neural Network Training through Dimensionality Reduction for High-Throughput Screening of Topological Materials}

\author{Ruman Moulik\footnote{phz218064@physics.iitd.ac.in}, Ankita Phutela, Sajjan Sheoran, Saswata Bhattacharya\footnote{saswata@physics.iitd.ac.in}} 
\affiliation{Department of Physics, Indian Institute of Technology Delhi, New Delhi 110016, India}

\begin{abstract}
	\noindent 
Machine Learning facilitates building a large variety of models, starting from elementary linear regression models to very complex neural networks. Neural networks are currently limited by the size of data provided and the huge computational cost of training a model. This is especially problematic when dealing with a large set of features without much prior knowledge of how good or bad each individual feature is. We try tackling the problem using dimensionality reduction algorithms to construct more meaningful features. We also compare the accuracy and training times of raw data and data transformed after dimensionality reduction to deduce a sufficient number of dimensions without sacrificing accuracy. The indicated estimation is done using a lighter decision tree-based algorithm, AdaBoost, as it trains faster than neural networks. We have chosen the data from an online database of topological materials, Materiae. Our final goal is to construct a model to predict the topological properties of new materials from elementary properties.
\end{abstract}
\pacs{}
\maketitle

\section{Introduction}
Topology is a modern branch of mathematics that emerged in the early 1900s \cite{aitkenreview}. It is a qualitative approach to geometry, to measure the properties of objects that remain unchanged when subjected to continuous deformation. It has been of great interest in the field of condensed matter physics \cite{kane2005z,bernevig2006quantum} in recent years, mainly because of the discovery of topological insulators (TI) \cite{fu2007topological,hasan2010colloquium,qi2011topological} and topological crystalline insulators (TCI) \cite{hsieh2012topological,benalcazar2017electric,kruthoff2017topological}. They are an entirely new class of quantum materials that act as an insulator in bulk but have conducting surface states. Those surface states have a Dirac-cone-like structure which facilitates dissipationless electronic currents \cite{diracintro,dirac}. This allows TIs and TCIs to have plenty of novel applications, inspiring the search for new materials of such kind with varying properties that suit different applications.

The surface states are protected by time-reversal symmetry in the case of TIs, and various crystal symmetries in the case of TCIs \cite{kitaev2009periodic,schnyder2008classification,chiu2016classification,schindler2018higher,song2017d}. Traditionally TIs have been characterised by their non-zero $Z_2$ invariant, while TCIs have been characterised by their non-zero Chern numbers \cite{moore2007topological}. Therefore, the prediction of topological materials requires tedious calculations of topological invariants or identification of topological nodes \cite{yu2011equivalent}. Such methods are limited to one material at a time \cite{phutela2022exploring}. Recently, the development of symmetry-based indicators \cite{symmetryIndicator} has enabled high-throughput screening for topological materials. Notably, the SymTopo \cite{symtopo} package was developed based on the above paradigm and has allowed high-throughput screening to search for new topologically non-trivial materials. The results of these calculations are available publicly in the Materiae \cite{materiae} database through the magic of application programming interface (API). However, looking at the Materials Project \cite{materialsProject} database, it is quickly realised that the number of unexplored materials is much greater than the ones explored for topological properties. The symmetry-based approaches only minimise the band structure calculations needed to predict topological behaviour but do not eliminate the need for such first-principles analyses. Such methods cannot hope to cover such a vast database in reasonable time and computational costs. The high resource cost calls for a predictive approach that would use previously calculated data, bringing us to Machine Learning (ML).

In recent years ML has taken the world by storm, finding a plethora of applications in every field possible like image and speech recognition \cite{zhang2019application,chen2021review,alharbi2021automatic}, weather and traffic predictions \cite{bochenek2022machine,shaygan2022traffic}, and medical diagnosis \cite{mirbabaie2021artificial,kumar2022artificial}. The basic idea of ML is to take pre-existing data and develop complex models to extrapolate them, a kind of fancy curve-fitting. There are various kinds of available predictive models, such as, regression \cite{rifkin2007notes}, support vector machines \cite{smola2004tutorial}, decision trees \cite{decisiontrees} and random forests \cite{breiman2001random} among which artificial neural networks (ANN) \cite{fukushima1975cognitron,feldman1982dynamic,ballard1987modular} are the most complex but also result in the best predictions. However, their substantial training expense poses a significant constraint which is noteworthy considering the large number of features we are aiming to train on. In this article, an attempt has been made to address the aforementioned issue using dimensionality reduction \cite{dimred} to streamline the number of features we use to train our model. Finally models are trained for predicting TIs and TCIs in various materials systems using ANN. The models are used to predict respective properties for materials taken from the Materials Project database. Electronic structure calculations are performed on the predicted materials to validate the accuracy of our models. 

\begin{figure}
	\centering
	\includegraphics[width=0.6\linewidth]{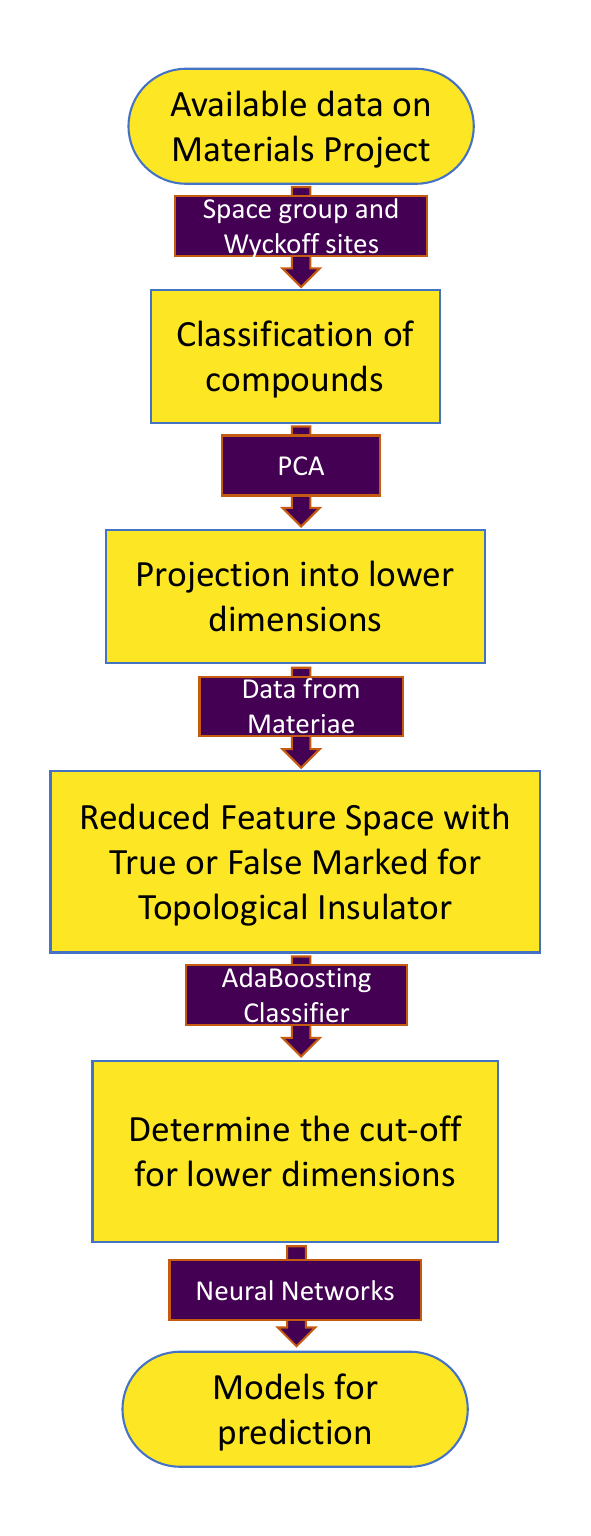}
	\caption{Flowchart of the methodology followed to construct predictive models using neural networks.}
	\label{fig:flowchart1}
\end{figure}

\section{Methodology \& Analysis}

\subsection{Data acquisition}
A flowchart of the methodology followed to construct predictive models using neural networks is illustrated in Fig. \ref{fig:flowchart1}. ML is a data-driven science, meaning a substantially sized topological property dataset is needed to construct a reliable prediction model. Topological properties can be calculated from \textit{ab initio} methods, viz., density functional theory (DFT) \cite{kohn1996density} taking spin-orbit coupling (SOC) into account and using various symmetry-based indicators \cite{symmetryIndicator}. These calculations have been performed on a lot of materials by researchers worldwide. However, the collection and access of the same can be cumbersome. Our work ahead is made more accessible by databases that archive such calculations and present them in an easily accessible format.

Materials Project \cite{materialsProject} is a growing database of materials containing structural data, electronic structures, various physical properties, etc., calculated using DFT. This database has been integrated with web-based technologies to allow the retrieval of data through API. A Python library has been developed to use the API easily and is available with the name mp-api. Materiae \cite{materiae} is a database dedicated to storing topological data. It uses structures from the Materials Project and an automated algorithm, SymTopo \cite{symtopo}, based on exhaustive mappings between symmetry representations of occupied bands and topological invariants. Their database also supports the use of APIs, and the standard Python requests library is used. As of \printdate{2023-02-15}, Materials Project had 154,718 materials, while Materiae had only 28,605 members in its catalogue. This size mismatch indicates how much of the available database is yet to be explored for topological properties.

\subsection{Site distinction}
As the intention is to train models using elemental properties as features, similar materials related by site substitutions are grouped together, and separate models are made for individual groups. For this purpose, we do a broad classification based on their space group (SG), number of sites ($N_s$), and number of distinct sites ($N_d$) of the compounds. To ensure that models are trained upon sizeable populations, the number of TIs is checked from Materiae for each group, and ones with sufficiently large numbers are chosen, as tabulated in Table \ref{table:TIcount}.
\begin{table}[]
	\caption{Number of TIs and TCIs in each chosen group of compounds.}
	\begin{tabular}{|m{0.07\textwidth}|m{0.07\textwidth}|m{0.12\textwidth}|}
\hline \textbf{SG} & \boldmath{$N_s$} & \raggedleft \textbf{No. of TIs and TCIs} \tabularnewline \hline 
62            & 12              & \raggedleft 129           \tabularnewline  \hline
139           & 5               & \raggedleft 130           \tabularnewline \hline
189           & 9               & \raggedleft 96            \tabularnewline \hline
225           & 2               & \raggedleft 64            \tabularnewline \hline
221           & 2               & \raggedleft 53            \tabularnewline \hline
221           & 5               & \raggedleft 43            \tabularnewline \hline
	\end{tabular}

	\label{table:TIcount}
\end{table}

For the sake of being stricter on the condition of substitutions, the sites are also checked by their Wyckoff positions, considering transformations between them. Atoms with the same Wyckoff positions across different materials are considered substitutions of each other. The different sites and considered Wyckoff schemes are tabulated in Table \ref{table:wyckoff}. However, binary systems ($N_d=2$) are treated differently as their sites can always be interconverted by suitable transformations. Such cases are distinguished by using electronegativities to order the respective elements. This concept is also extended to systems with more than two sites but multiple sites having the same Wyckoff positions, like the system with SG = 12;$N_s=12$. Compounds with a different Wyckoff position scheme from the majority are discarded.

\begin{table}[]
	\caption{Wyckoff indices being considered as direct substitutions in lattice for materials with 3 distinct sites per unit cell.}
	\begin{tabular}{m{0.05\textwidth}m{0.05\textwidth}m{0.05\textwidth}m{0.07\textwidth}m{0.07\textwidth}m{0.07\textwidth}}
		\hline
		\multicolumn{1}{|l|}{\multirow{2}{*}{\textbf{SG}}} & \multicolumn{1}{l|}{\multirow{2}{*}{\boldmath $N_s$}} & \multicolumn{1}{l|}{\multirow{2}{*}{\boldmath $N_d$}} & \multicolumn{3}{l|}{ \textbf{Wyckoff Positions}}                                                                                      \tabularnewline \cline{4-6} 
		\multicolumn{1}{|l|}{}                             & \multicolumn{1}{l|}{}                                 & \multicolumn{1}{l|}{}                                          & \multicolumn{1}{l|}{\textbf{Site 1}}             & \multicolumn{1}{l|}{\textbf{Site 2}}             & \multicolumn{1}{l|}{\textbf{Site 3}}             \tabularnewline \hline
		\multicolumn{1}{|l|}{\multirow{2}{*}{62}}          & \multicolumn{1}{l|}{\multirow{2}{*}{12}}              & \multicolumn{1}{l|}{\multirow{2}{*}{3}}                        & \multicolumn{1}{l|}{\multirow{2}{*}{c}} & \multicolumn{1}{l|}{\multirow{2}{*}{c}} & \multicolumn{1}{l|}{\multirow{2}{*}{c}} \tabularnewline
		\multicolumn{1}{|l|}{}                             & \multicolumn{1}{l|}{}                                 & \multicolumn{1}{l|}{}                                          & \multicolumn{1}{l|}{}                   & \multicolumn{1}{l|}{}                   & \multicolumn{1}{l|}{}                   \tabularnewline \hline
		\multicolumn{1}{|l|}{\multirow{2}{*}{139}}         & \multicolumn{1}{l|}{\multirow{2}{*}{5}}               & \multicolumn{1}{l|}{\multirow{2}{*}{3}}                        & \multicolumn{1}{l|}{a}                  & \multicolumn{1}{l|}{d}                  & \multicolumn{1}{l|}{e}                  \tabularnewline \cline{4-6} 
		\multicolumn{1}{|l|}{}                             & \multicolumn{1}{l|}{}                                 & \multicolumn{1}{l|}{}                                          & \multicolumn{1}{l|}{b}                  & \multicolumn{1}{l|}{d}                  & \multicolumn{1}{l|}{e}                  \tabularnewline \hline
		\multicolumn{1}{|l|}{\multirow{2}{*}{189}}         & \multicolumn{1}{l|}{\multirow{2}{*}{9}}               & \multicolumn{1}{l|}{\multirow{2}{*}{3}}                        & \multicolumn{1}{l|}{a}                  & \multicolumn{1}{l|}{f}                  & \multicolumn{1}{l|}{g}                  \tabularnewline \cline{4-6} 
		\multicolumn{1}{|l|}{}                             & \multicolumn{1}{l|}{}                                 & \multicolumn{1}{l|}{}                                          & \multicolumn{1}{l|}{b}                  & \multicolumn{1}{l|}{g}                  & \multicolumn{1}{l|}{f}                  \tabularnewline \hline
		\multicolumn{1}{|l|}{\multirow{2}{*}{221}}         & \multicolumn{1}{l|}{\multirow{2}{*}{5}}               & \multicolumn{1}{l|}{\multirow{2}{*}{3}}                        & \multicolumn{1}{l|}{a}                  & \multicolumn{1}{l|}{b}                  & \multicolumn{1}{l|}{c}                  \tabularnewline \cline{4-6} 
		\multicolumn{1}{|l|}{}                             & \multicolumn{1}{l|}{}                                 & \multicolumn{1}{l|}{}                                          & \multicolumn{1}{l|}{b}                  & \multicolumn{1}{l|}{a}                  & \multicolumn{1}{l|}{d}                  \tabularnewline \hline
		&                                                       &                                                                &                                         &                                         &                                        
	\end{tabular}
	\label{table:wyckoff}
\end{table}

\subsection{Construction of feature space}
Previous attempts to predict properties of materials from elementary properties of atoms provide insight into constructing our feature space \cite{Matynov-Batsanovelectronegativity,ouyang2019exploiting,cao2020artificial,wan2021machine}. The idea is to build an ample feature space to accommodate as many varied properties as possible and then let various ML methods choose the significant ones for our data. A total of 12 features for each atom are selected from our survey. They are electronegativity, atomic number, ionisation energy, electron affinity, atomic radius, number of valence electrons, Matynov-Batsanov electronegativity \cite{Matynov-Batsanovelectronegativity}, pseudopotential core radius, and number of electrons in the outermost \textit{s}-, \textit{p}-, \textit{d}- and \textit{f}-orbitals. This translates to 24 features for $N_d=2$ and 36 features for $N_d=3$. The problem we are addressing is a classification problem of topologically active materials. We consider those classified on Materiae as TI or TCI as ``true" and others as ``false", and our predictions follow the same classification.

\subsection{Normalisation and dimensionality reduction}
The ideal data distribution is usually a normal curve. Most predictive models have better accuracy when applied to normal distributions. A power transformer transforms the data to have zero-mean, unit-variance satisfying the condition. The Yeo-Johnson method \cite{yeo2000} implemented in the scikit-learn \cite{scikit} library is used for this step.

As discussed earlier, a total of 12 features for each site are considered, i.e., 24 features for binary materials and 36 features for those with three $N_d$. Such a considerably large number of features makes it hard to train a neural network in a reasonable time. Choosing to train on such a vast number of features also has the potential to confuse the final model if too many non-correlated features are provided. For this reason, a principal component analysis (PCA) \cite{pca,pca2} is performed to take better projections of features that correlate better with the data and reduce the number of features to build our final model.

The scikit-learn package provides different PCA methods, using different kernels like linear, polynomial, exponential, etc. Kernels help project the linearly inseparable data onto a higher dimensional space where they might be linearly separable. The implementation of PCA with kernels is called kernel principal component analysis (KPCA) \cite{kernelpca}. The various kernels are used, and results are plotted up to three dimensions to check which method gives better visual separation of the data, which are included in the supplemental material (SM). It is observed that using KPCA with radial basis function kernel results in the best visual separation. The data is transformed into its principal components in this step, which is used for further studies.

\subsection{Dimension Pruning}
While PCA can construct the best dimensions for use, it provides no useful information for determining the number of dimensions to be chosen. For this purpose, a decision tree-based model \cite{decisiontrees} is trained on both the original data and the data transformed from PCA but with varying dimensions. The reason to choose a decision tree-based model is their comparatively faster training time while maintaining the ability to build complex models. Adaboost \cite{druckeradaboost,adaboost} is a good choice for our binary classification problem, with the advantage of having relatively fewer hyperparameters as well as lesser chances of overfitting.
\begin{figure*}[t]
	\centering
	\includegraphics[ ]{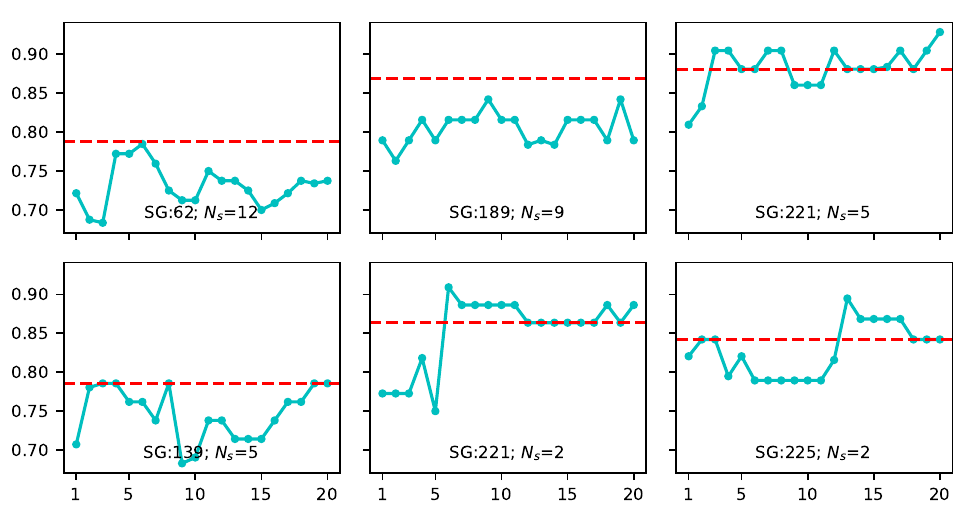}
	\caption{Comparison of accuracies of training models using raw data and increasing number of dimensions from PCA. Red dotted lines represent the raw data while the cyan lines represent the respective dimensions, as given on the x-axis. The accuracies are given on the y-axis.}
	\label{fig:adaboost}
\end{figure*}

\begin{figure}[t]
	\centering
	\includegraphics[]{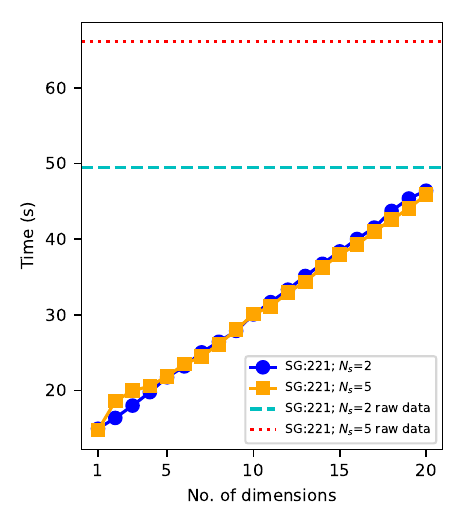}
	\caption{Comparison of training times of adaboost on various number of dimensions. Only SG = 221 is shown, but other datasets show the same patterns.}
	\label{fig:adaboosttimes}
\end{figure}
The PCA-transformed dataset of each system is used to train Adaboost models, starting from a single dimension to twenty dimensions. Using the same data to learn the parameters of a prediction model and testing the model is usually considered a mistake. This is because we can not be confident about the predictions on yet unknown data. Hence, some of the available training data is actually held out of the training set to make a test set to be used for validation. A common practice is to do a $k$-fold cross-validation \cite{breimancv,kohavicv} in which the data is equally split into $k$ parts. A model is trained on $k-1$ parts, and remaining 1 part is used to score the model. $k$ different models with respective scores are obtained by changing the part of the data not used in training. Finally, the model with the highest score is chosen. Five-fold cross-validation is used at each step to help select better models. The accuracy of each model vs. the number of dimensions being considered is plotted, as in Fig. \ref{fig:adaboost}. The two hyperparameters, viz., learning rate and the maximum number of trees, are tuned such that further increasing them doesn't change the nature of the plot mentioned above. This is observed at a low learning rate of 0.01 and maximum estimators of 2500 and above. The same model is also trained on the raw data to compare accuracy and training times. Fig. \ref{fig:adaboosttimes} shows the respective plots and a suitable number of dimensions ($dim$) are chosen to achieve the best trade-off of accuracy vs. computational times, as given in Table \ref{table:hyperparameters}.
It is observed that about 3-6 dimensions are sufficient to achieve almost the accuracy of the raw data in most cases. The time comparison shows a linear increase with $dim$. Choosing $dim=5$ improves training time by about two-fold for systems with $N_d=2$, i.e., 24 features, and nearly three-fold for systems with $N_d=3$, i.e., 36 features.

\subsection{Neural Networks for Prediction}
Neural networks are chosen as the final model we train to output predictions of topological materials due to their generally high complexity and accuracy. Their primary disadvantage is the model training cost, but once trained, the predictions can be churned out quickly and en masse if needed. A Multi-Layer Perceptron (MLP) \cite{hinton1990connectionist,fakhr1990fast} classifier is used, as implemented in the scikit-learn library. The default setting of using Adam solver \cite{adam} is not changed. In the case of MLP, unlike AdaBoost, a lot of hyperparameters are to be considered, i.e., learning rate ($lr$), strength of the L2 regularization ($\alpha$), number of hidden layers ($hl_d$), sizes of hidden layers ($hl_s$), and maximum number of iterations ($n_i$) to run for. These have been extensively tested across multiple values using the Optuna framework \cite{optuna_2019} and the best ones were chosen for each dataset individually. A five-fold cross-validation method is also implemented within the scoring model to help select a better predictive model. It is in this step that our earlier method of reducing the overall time taken to train a neural network becomes significant. The computational cost of extensively testing models across a wide range of hyperparameters along with cross-validation would be very high. The achieved validation scores and corresponding hyperparameters are listed in Table \ref{table:hyperparameters}.
\begin{table}[]
	\caption{Summary of hyperparameters used in training our MLPs. It lists the number of dimensions kept from PCA and all the parameters passed as arguments while training our final model}
	\begin{tabular}{|>{\centering\arraybackslash}m{0.07\textwidth}|>{\centering\arraybackslash}m{0.04\textwidth}|>{\centering\arraybackslash}m{0.05\textwidth}|>{\centering\arraybackslash}m{0.1\textwidth}|>{\centering\arraybackslash}m{0.1\textwidth}|>{\centering\arraybackslash}m{0.03\textwidth}|>{\centering\arraybackslash}m{0.03\textwidth}|}
		\hline
		\textbf{SG}/\boldmath{$N_S$} & \boldmath{$dim$} & \boldmath{$n_i$} & \boldmath{$lr$}      & \boldmath{$\alpha$}  & \boldmath{$hl_s$} & \boldmath{$hl_d$} \\ \hline
		62/12  & 6   & 2000  & 1.5$\times10^{-4}$  & 1.347$\times10^{-4}$  & 135    & 3      \\ \hline
		139/5  & 4   & 1108  & 1.86$\times10^{-3}$ & 2.3851$\times10^{-5}$ & 93     & 9      \\ \hline
		189/9  & 4   & 2234  & 3.19$\times10^{-3}$ & 2.486$\times10^{-5}$  & 146    & 9      \\ \hline
		221/2  & 6   & 3684  & 1.48$\times10^{-4}$ & 1.197$\times10^{-5}$  & 73     & 7      \\ \hline
		221/5  & 3   & 2800  & 6$\times10^{-3}$   & 3.5$\times10^{-5}$    & 75     & 10     \\ \hline
		225/2  & 3   & 3999  & 6.99$\times10^{-3}$ & 3.447$\times10^{-5}$  & 79     & 10     \\ \hline
	\end{tabular}
	\label{table:hyperparameters}
\end{table}

\begin{figure}
	\centering
	\includegraphics[width=0.6\linewidth]{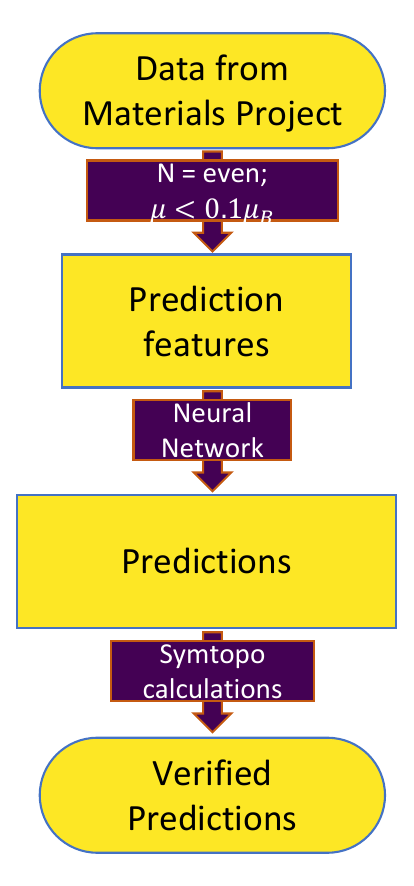}
	\caption[]{The flowchart of the methodology followed for predictions from our models.}
	\label{fig:flowchart2}
\end{figure}

To validate our predictions, we have adopted the same method as the Materiae website, which uses the SymTopo package. It is to be kept in mind that their theory only works for nonmagnetic materials. Additionally it performs a hard-check on number of electrons in the DFT calculation, only accepting an even number of electrons. This property is masked by our methodology due to the initial transformation and PCA of our original features. Materials that are not included in Materiae are picked up from Materials Project and checked for the former two conditions. Features are generated for the ones that pass. Previously trained power transformer, PCA transformation and MLP prediction are applied in that order. Positive outcomes from our predictions are then tested using SymTopo for further validation and reported. The steps followed for predictions are represented as a flowchart in Fig. \ref{fig:flowchart2}. SymTopo is used to prepare input files for Vienna ab initio simulation package (VASP) \cite{kresse1993vasp,kresse1996efficiency,kresse1996efficient}, which performs the actual electronic structure calculations according to DFT with projector augmented wave (PAW) potentials \cite{kresse1999ultrasoft}. Perdew–Burke-Ernzerhof (PBE) \cite{perdew1996generalized} exchange correlation function is used. The results from SymTopo are discussed in the section below.

\section{Summary of Predictions}

The count of different kind of materials as output from SymTopo are listed in Table \ref{table:results}. The number in ``Total" column only includes candidates with an even number of valence electrons, nonmagnetic nature, i.e., very low magnetic moment and predicted as "true" for non-trivial topological behaviour by our models. The correct predictions made by our model are those which SymTopo also computes as being of the TI or TCI category. The ``Semimetal" column includes both high-symmetry point semimetals and high-symmetry line semimetals. Both systems are topologically non-trivial with topological nodes at high-symmetry points or lines respectively \cite{SymTopo}. However, they are of less interest to us as they cannot be used as TI or TCI due to their degeneracy at high-symmetry points or lines, and were consequently marked as false in our training data. However, due to their non-trivial topological nature, a lot of predicted TIs from our model are found to be these semi-metals instead but they cannot be strictly classified as false positive results. The ``Trivial" column consists of topologically trivial insulators and are strictly false positives in the case of our predictions.

Table \ref{table:results} illustrates that certain material systems are ripe for exploration of topological properties, particularly the first three that have been chosen. The predictions of systems with SG = 12;$N_s=12$, SG = 139;$N_s=5$ and SG = 189;$N_s=9$ have 60\%, 50\% and 63\% correctly predicted TI and TCI, respectively, along with a very low percentage of trivial insulators. In those cases it is clear that our method has yielded a high percentage of topologically active materials with no electronic structure calculations of our own, proving the potential of similar data-based studies for exploration of topological properties. The results are not that spectacular for the other three systems, which means our models are not able to capture the essence from their data, or maybe those systems do not have too much left to explore for non-trivial topology. As more data becomes available, the method can be extended to more systems. Material-wise detailed catalogue of classification has been made available in the SM. 

\begin{table}[]
	\caption{Summary of predictions verified by SymTopo}
	\begin{tabular}{|>{\centering\arraybackslash}m{0.07\textwidth}|>{\centering\arraybackslash}m{0.12\textwidth}|>{\centering\arraybackslash}m{0.05\textwidth}|>{\centering\arraybackslash}m{0.05\textwidth}|>{\centering\arraybackslash}m{0.06\textwidth}|>{\centering\arraybackslash}m{0.05\textwidth}|}
		\hline
		\textbf{SG}/\boldmath{$N_S$} & \textbf{Semimetal} & \textbf{TI} & \textbf{TCI} & \textbf{Trivial} & \textbf{Total} \\ \hline
		62/12   & 25         & 36 & 6   & 3       & 70    \\ \hline
		139/5   & 4          & 4  & 3   & 3       & 14    \\ \hline
		189/9   & 10         & 2  & 15  & 0       & 27    \\ \hline
		221/2   & 6          & 0  & 0   & 6       & 12    \\ \hline
		221/5   & 13         & 1  & 1   & 13      & 28    \\ \hline
		225/2   & 0          & 0  & 1   & 1       & 2     \\ \hline
	\end{tabular}	
	\label{table:results}
\end{table}

\section{Conclusion}

To conclude, we have construced different MLP models with various accuracies calculated over a five-fold cross-validation method. In the process, we have demonstrated the viability of using dimensionality reduction to reduce the number of features used to train the neural networks, significantly reducing the training times. We have also proposed a method using decision tree-based models to help select the number of dimensions that should be chosen from the PCA analysis. All of this now allows us to make predictions about topological properties of new proposed materials related to our chosen systems by site substitutions much faster and to scan a much larger materials space with very little computation. The predictions have been verified using the algorithm used to generate the training data in the first place. We expect the method to extend to more systems as more data becomes available. Furthermore, the use of dimensionality reduction to reduce training times appears to be an independent property that might be extensible to more ML applications beyond the particular application presented in this paper.

\section{Acknowledgement}
R.M. acknowledges CSIR, India, for the junior research fellowship [Grant No. 09/0086(12865)/2021-EMR-I]. A.P. acknowledges IIT Delhi for the senior research fellowship. S.S. acknowledges CSIR, India,
for the senior research fellowship [grant no. 09/086(1432)/2019-EMR-I]. S.B. acknowledges financial support from SERB under a core research grant (Grant No. CRG/2019/000647) to set up his high-performance computing (HPC) facility “Veena” at IIT Delhi for computational
resources.

\bibliography{ref}

\end{document}